\documentclass[prd,preprint,tightenlines,floatfix,showpacs,preprintnumbers,nofootinbib,eqsecnum]{revtex4}
\usepackage[dvips,final]{graphicx}
  \usepackage{amssymb}
   \usepackage{amsmath}
    \usepackage{amsfonts}
     \usepackage{epsfig}
      \usepackage{bm}% bold math
\renewcommand{\epsilon}{\varepsilon}
\begin{document}

\thispagestyle{empty} \preprint{\hbox{}} \vspace*{-10mm}

\title{Finite-width effects in the mixing of neutral mesons}

\author{V.~I.~Kuksa}

\email{kuksa@list.ru}

\affiliation{Institute of Physics,
Southern Federal University ,
Rostov-on-Don 344090, Russia}

\date{\today}

\begin{abstract}
We analyze a contribution of the finite-width (mass-smearing) effects to the mixing of neutral mesons. It was shown, that this contribution is dominant in the $D$-meson system and large in the $K$-meson one. An account of the mass-smearing effects allows to explain some discrepancy between standard predictions and experimental data in these cases.
\end{abstract}

\pacs{11.30.Er,11.30.Pb}

\maketitle

%---------------------
\section{Introduction}

Finite-width effects (FWE) is an analog of the energy level spreading which takes place in the quantum non-stationary systems. These effects arise in the processes with participation of the unstable particles (UP) with a large width. There are two principal ways of description of FWE (or instability). In the $S$-matrix approach UP in an intermediate state is described by the amplitude with complex pole \cite{1} or by dressed propagator \cite{2}. The second way is the description of instability by the time-dependent field operator \cite{3,4} which simulates an "asymptotic" state of UP.

It was shown in Ref.~\cite{3}, that the instability is connected with the spreading (smearing) of the UP mass. A wave function of UP in their rest system can be written in terms of its Fourier transform \cite{3}:
\begin{equation}\label{1.1}
 \Psi(t)=exp\{iMt-\Gamma |t|/2\} \rightarrow \frac{\Gamma}{2\pi}\int \frac{exp\{-imt\}}{(m-M)^2+\Gamma^2/4}\,dm.
\end{equation}
Right-hand part of Eq.~(\ref{1.1}) may be interpreted as a distribution of mass value $m$, with a spread $\delta m$, related to the lifetime $\delta \tau =1/\Gamma$, by an uncertainty relation $\delta m \cdot \delta \tau \sim 1$.

Uncertainty relations play a fundamental role in quantum theory. They are based on the general principles of the theory and manifest themselves in the processes on various hierarchy levels. There are two different types of uncertainty relations in quantum theory: the Heisenberg uncertainty relations and the time-energy uncertainty relation \cite{4a}. The first type relation takes place for canonically conjugated quantities and follows from the Cauchy-Schwarz inequality and commutation relations for operators, corresponding to these quantities \cite{4b}. The standard Heisenberg uncertainty relation is valid for the operators of the momentum and coordinate. There is no any operator which corresponds to the time and the second type uncertainty relation has a completely different character. The time-energy uncertainty relation follows from the equation of motion in the Heisenberg representation, which desribes the evolution of the non-stationary quantum system \cite{4c,4d}. In this case, the uncertainty relation can be represented in the form \cite{4a}:
\begin{equation}\label{1.2}
\Delta E\cdot\Delta t\geq\frac{1}{2},\,\mbox{where}\,\Delta t=\frac{\Delta O(t)}
{|\frac{d}{dt}\bar{O}(t)|}.
\end{equation}
Thus, $\Delta t$ is the time interval, during which the physical value, described by operator $O(t)$, undergos to the character variation. In the case of unstable particle, $\Delta t$ is the lifetime and $\Delta E$ is the value of the mass smearing $\Delta m$ in the rest frame system .

The effect of the mass smearing (or finite-width effect) was applied in Refs.~\cite{5,6} for the description of UP with smeared mass by the wave field function in the most general form:
\begin{equation}\label{1.3}
 \Psi(x)=\int\Psi(x,\mu)\omega(\mu)d\mu\,.
\end{equation}
Here $\Psi(x,\mu)$ is the spectral component of wave function which describes particle with fixed mass squared $m^2=\mu$ and $\omega(\mu)$ is some weight function. The function $\rho(\mu)=|\omega(\mu)|^2$ describes the distribution (smearing, spreading) of random (fuzzed) mass $m$ of UP. This smearing is caused by the stochastic self-energy type interaction of UP with the vacuum fluctuations \cite{6} and leads to FWE. There are many processes with the participation of UP, where FWE play a significant role \cite{6}.

In this work, we consider the contribution of  FWE or mass smearing to the mass difference in the neutral meson systems $M^0-\bar{M}^0$. Mass difference $\Delta m=m_H-m_L$ of heavy ($M^0_H$) and light ($M^0_L$) component is one of the main characteristics of mixing in $M^0-\bar{M}^0$ system, which describes the time of oscillations. When $\Delta m \sim \Gamma_S$, i.e. mass difference is an order of width of the short-lived component $M^0_S$, then the smearing of mass of this component leads to an additional contribution to the total mass difference. To illustrate the situation we present in Table 1 the experimental data  on $\Delta m$ and $\Gamma$ for $K$, $B$ and $D$ mesons, which are taken from Refs.\cite{7,8,9}.

\begin{center}
 \begin{tabular}{||c|c|c|c||}
  \hline
  $M^0-\bar{M}^0$     & $\Delta m^{exp}$ (eV)             & $\Gamma^{exp}$ (eV)         \\ \hline
  $K^0-\bar{K}^0$     & $(3.483\pm 0.006)\cdot 10^{-6}$   & $7.349\cdot 10^{-6}$  \\ \hline
  $B^0_d-\bar{B}^0_d$ & $(3.337\pm 0.033)\cdot 10^{-4}$   & $4.301\cdot 10^{-4}$  \\ \hline
  $B^0_s-\bar{B}^0_s$ & $(117\pm0.8)\cdot 10^{-4}$      & $4.488\cdot 10^{-4}$  \\ \hline
  $D^0-\bar{D}^0$     & $(1.56\pm 0.5)\cdot 10^{-5}$      & $1.60\cdot 10^{-3}$    \\ \hline
 \end{tabular}
\end{center}
From the data in Table 1, it follows that $\Delta m_K \approx \Gamma_K/2$, $\Delta m_{B_d} \leq \Gamma_{B_d}$ and $\Delta m_D \ll \Gamma_D$. So, the contribution of the mass-smearing effect (MSE ) to the mass difference can be large for the $K$ and $D$ mesons and noticeable for $B_d$ meson. In the case of $B_s$ meson, this contribution is negligible due to $\Gamma\ll \Delta m$.

The aim of this work is to evaluate the contribution of the MSE to the mass difference for the case of $K$, $B_d$ and $D$ mesons. In the second section, we consider the theoretical status of mixing in the standard approach. The formalism for evaluation of mass-smearing contribution to the mixing in the general case is presented in Section III. The evaluation of this contribution to the mixing in $K^0-\bar{K}^0$, $B^0_d-\bar{B}^0_d$ and $D^0-\bar{D}^0$ systems is fulfilled in Sections IV and V. It was shown, that the contribution of mass smearing to mixing in $D^0$-meson and $K^0$-meson is dominant and large respectively.

\section{Mass difference in the Standard Model}

In the standard approach mass difference is defined by the relation $\Delta m \approx 2 Re(M_{12})$, where $M_{12}$ is the amplitude of the flavor changing transition $M^0\to \bar{M}^0$. In the framework of Standard Model it is caused by box diagrams which lead to the flavor changing neutral current (FCNC) at one-loop level (short-distance contribution, SD) \cite{10}. Long-distance contribution is uncontrolled in the case of $K$ and $D$ mesons and is assumed negligible in the mixing of $B$ mesons.

The SD value of mass difference in the case of $K^0-\bar{K}^0$ systems is \cite{10,11}:
\begin{equation}\label{2.1}
 \Delta m_K=\frac{G^2_F}{6\pi^2}M^2_W M_K B_K f^2_K [\eta_1 \lambda^2_c s_0(x_c)+
            \lambda^2_t \eta_2 s_t (x_t)+2\lambda_c \lambda_t \eta_3 s_0(x_c ,x_t)].
\end{equation}
In Eq.(\ref{2.1}) the notations are in general usage:
\begin{align}\label{2.2}
 &\lambda^2_c=|U^*_{cs}U_{cd}|^2,\,\,\,\lambda^2_t=|U^*_{ts}U_{td}|^2,\,\,\,\mbox{(U is KM matrix)}, \notag\\
 &\lambda_c\lambda_t=|U^*_{cs}U_{cd}|\cdot |U^*_{ts}U_{td}|,\,\,\,s_0(x_c)=x_c=\frac{m^2_c}{M^2_W},\notag\\
 &s_0(x_t)=\frac{4x_t-11x^2_t+x^3_t}{4(1-x_t)^2}-\frac{3x^3_t}{2(1-x_t)^3}\ln {x_t},\,\,\,x_t=\frac{m^2_t}{M^2_W},\notag\\
 &s_0(x_c,x_t)=x_c [\ln {\frac{x_t}{x_c}}-\frac{3x_t}{2(1-x_t)}-\frac{3x^2_t}{4(1-x_t)^2}\ln {x_t}].
\end{align}
The values of parameters, entering into Eq.(\ref{2.1}), are discussed in Refs.\cite{10,11}:
\begin{align}\label{2.3}
 &B_K=0.84,\,\,\,\eta_1=1.38,\,\,\,\eta_2=0.574,\,\,\,\eta_3=0.47,\,\,\,f_K=160 \,\mbox{MeV}\,\,\, [14];\notag\\
 &B_K=0.86\pm0.15,\,\,\,\eta_1=1.38\pm0.20,\,\,\,\eta_2=0.57,\,\,\,\eta_3=0.47,
 \,\,\,f_K=160\, \mbox{MeV}\,\,\, [15].
\end{align}
 The numerous lattice evaluations of $B_K$ give significantly less values, for instance, $B^{RI}_K(2\,\mbox{GeV})=0.514$, $B^{\overline{MS}}_K (2\,\mbox{GeV})=0.524$ and $\hat{B}_K=0.72$ \cite{12}. A recent review, including all lattice data, quoted a value $B^{\overline{MS}}_K (2\,\mbox{GeV})=0.58(3)(6)$ \cite{12a}. From Eqs.(\ref{2.1})-(\ref{2.3}) one can get $\Delta m_K\approx 2.34\cdot 10^{-6}\,\mbox{eV}$. We consider also the result, given in Ref.\cite{13}, where lattice $B_K$ was used:
\begin{equation}\label{2.4}
 \Delta m_K=(1.87\pm 0.49)\cdot 10^{-6}\, \mbox{eV}.
\end{equation}
 So, the standard evaluations of SD contribution to mass difference give the values which are significantly less than the experimental one:
\begin{equation}\label{2.5}
 \Delta m^{exp}_K =(3.483\pm0.006)\cdot 10^{-6}\,\mbox{eV}.
\end{equation}
This discrepancy has been the subject of various speculations concern extra FCNC transition: unknown LD contribution, mixing of singlet quark with ordinary ones \cite{14}, the presence of additional $Z^{'}$ boson \cite{15}, charged Higgs boson \cite{16} and other new physics \cite{16a}. In Section IV, we show that an account of the MSE can significantly increases the theoretical mass difference and improve the correspondence of theoretical predictions  within the framework of Standard Model (SM) and experimental data .

Mass difference in $B^0_d-\bar{B}^0_d$ system is mainly caused by $t$-quark contribution \cite{10}:
\begin{equation}\label{2.6}
 \Delta m_B=\frac{G^2_F}{6\pi^2}M^2_W M_B B_B f^2_B\eta_B s_0(x_t)|U^*_{td}U_{tb}|^2,
\end{equation}
where for $x_0(x_t)$ a good approximation $x_0(x_t)=0.784x^{0.76}_t$ is given in \cite{16b}. The values of parameters are discussed in Refs.\cite{10}, \cite{11}, \cite{17} and \cite{18}:
\begin{align}\label{2.7}
 &\eta_B=0.551\,\,\,[14],\,\,\,\eta_B=0.55\,\,\,[15];\notag\\
 &f_B=0.20\pm0.03\,\,\,\mbox{Gev}\,\,\,[15,24],\,\,\,B_B=1.30\pm0.12
 \,\,\,[15,24];\notag\\
 &f_B\sqrt{B_B}=0.228\,\,\,\mbox{GeV}\,\,\,[15,24],\,\,\,f_B\sqrt{B_B}=0.220
 \,\,\,\mbox{GeV}\,\,\,[25].
\end{align}
Here, we consider the results, given in Refs.\cite{17,18}:
\begin{equation}\label{2.8}
 \Delta m_B=3.646\cdot 10^{-4}\,\mbox{eV}\,\,\,[24],\,\,\,\Delta m_B=3.518\cdot 10^{-4}\,\mbox{eV}\,\,\,[25].
\end{equation}
The theoretical predictions (\ref{2.8}) are calculated for the average values of input parameters and exceed the experimental date:
\begin{equation}\label{2.9}
 \Delta m^{exp}_B=(3.337\pm 0.033)\cdot 10^{-4}\,\mbox{eV}.
\end{equation}
The deviation of theoretical values from experimental one is not large and can be explained by an uncertainty of the values $f_B\sqrt{B_B}$ and $|U^*_{td}U_{tb}|$. In Section V we show, that an account of the MSE decreases this deviation and improves correspondence between mean theory value and experimental data.

The mixing in $B^0_s-\bar{B}^0_s$ system is intensively studied now both in theoretical and experimental directions. However, as was noted in the first section, mass-smearing contribution to the mixing in this system is most likely very small.

In the case of $D^0-\bar{D}^0$ system, the situation is more complicated and hazy. The central problem of calculations is that two approaches that have been successful in treating heavy ($B$) and light ($K$) meson mixing both are not applicable to $D$ mixing \cite{19}. Moreover, due to relatively small splitting of down quarks GIM mechanism leads to strong damping of box contribution. Including box diagrams only, one finds $x^{box}\sim 10^{-5}$ and $y^{box}\sim 10^{-7}$ \cite{20}, where $x=\Delta m/\Gamma$ and $y=\Delta \Gamma/2\Gamma$. In the approaches, which are based on operator product expansion, the calculations usually yield $x,y\leq 10^{-3}$. The collaborations BaBar and Belle now obtained evidence for a non-vanishing mixing in the $D$ system: $x_D\sim 10^{-2}$ and $y_D\sim 10^{-3}$ \cite{9} (see Table 1). So, the experimental mass difference in the case of $D$ meson exceeds an expected theoretical one at least an order of magnitude (see also \cite{21}-\cite{24}). In Section V, we show that the contribution of mass-smearing to $x_D$ can be large and dominant. An account of mass smearing makes it possible to get an accordance between the theoretical value $x_D$ and experimental one.

\section{Description of the mass-smearing effect in $M^0-\bar{M}^0$ systems}

The mass difference is one of the main characteristics of the mixing and oscillation in the $M^0-\bar{M}^0$ systems. This mixing is caused by FCNC transition $M^0\longleftrightarrow \bar{M}^0$, which is described by box diagrams (SD contribution) within the frame-work of the Standard Model. The effect of mixing leads to the processes of type $M^0(t)\to \bar{M}^0(t)\to \bar{F}$ along with the straight process $M^0\to F$, where $F$ is flavor-specific final state. The ratio of the yields $Y(F)$ and $Y(\bar{F})$ is related with the mixing and can be observed as time-dependent oscillation of quark flavor.

In the frame-work of the model \cite{5}, the MSE is described by the probability density of smeared mass $\rho(m)$. In the case of $K^0$ mesons $\Delta m \sim \Gamma_S \gg \Gamma_L$, that is the short-lived component contributes to the additional mixing, and we need the functions $\rho(m_S)$. In the cases of $B^0$ and $D^0$ mesons, $\Gamma_S \sim \Gamma_L$ and both components contribute to the mixing. In the general case, we have $\overline{\Delta m}=\Delta m_0=M_H-M_L$, but it does not mean that the additional contribution of the mass smearing to the mass difference is equal to zero. Mass difference is not a directly observed physical value, nor the value we should average. Such a quantity is the transition probability $P(M^0\to \bar{M}^0(t))$, which is directly related with the mixing as function of $\Delta m$. Now we give a brief description of the conventional formalism we need for the calculations of mass-smearing contribution to mixing.

Time-dependent amplitudes of the transitions
\begin{equation}\label{3.1}
 \mathit{M}_{+}(t)=<M^0|M^0(t)>,\,\,\,\mathit{M}_{-}(t)=<\bar{M}^0|M^0(t)>
\end{equation}
in the parametrization of states \cite{16b}
\begin{align}\label{3.2}
  &|M^0(t)>=g_{+}(t)|M^0>+\frac{q}{p}g_{-}(t)|\bar{M}^0>,\notag\\
  &|\bar{M}^0(t)>=g_{+}(t)|\bar{M}^0>+\frac{p}{q}g_{-}(t)|M^0>
\end{align}
can be represented in the forms:
\begin{equation}\label{3.3}
 \mathit{M}_{+}(t)=g_{+}(t),\,\,\,\mathit{M}_{-}(t)=\frac{q}{p}g_{-}(t),\,\,\,
 \frac{q}{p}=(\frac{M^{*}_{12}-\frac{i}{2}\Gamma^{*}_{12}}{M_{12}-\frac{i}{2}\Gamma_{12}})^{1/2}.
\end{equation}
Then, the probabilities of the transitions are:
\begin{equation}\label{3.4}
 P_{+}(t)=|\mathit{M}_{+}(t)|^2=|g_{+}(t)|^2,\,\,\,P_{-}(t)=|\mathit{M}_{-}(t)|^2=
 |\frac{q}{p}|^2|g_{-}(t)|^2,
\end{equation}
where \cite{16b}:
\begin{equation}\label{3.5}
 |g_{\pm}(t)|^2=\frac{1}{2}\,e^{-\Gamma t}\,[\cosh(\frac{\Delta\Gamma}{2}t)\pm\cos(\Delta m\cdot t)].
\end{equation}
In Eq.(\ref{3.5}), $\Gamma=(\Gamma_{S}+\Gamma_{L})/2$, where $\Gamma_{S,L}$ are the widths of short- and long-lived (or heavy and light) states. Time-integrated probabilities are the function of $\Delta m$:
\begin{equation}\label{3.6}
 P_{\pm}(\Delta m)=\int P_{\pm}(t)\,dt.
\end{equation}
Using Eq.(\ref{3.5}), from (\ref{3.6}) one can get ($\Delta m\to x=\Delta m/\Gamma$):
\begin{equation}\label{3.7}
 P_{+}(x)=\frac{1}{2\Gamma}\,\{\frac{1}{1-y^2}+\frac{1}{1+x^2}\},\,\,\,
 P_{-}(x)=\frac{1}{2\Gamma}\,|\frac{q}{p}|^2\,\{\frac{1}{1-y^2}-\frac{1}{1+x^2}\},
\end{equation}
where $x=\Delta m/\Gamma$ and $y=\Delta\Gamma/2\Gamma$.

To account the mass-smearing effect, we have to average $P_{\pm}(x)=|M_{\pm}(x)|^2$ with the help of the model probability density $\rho_x(x)$ \cite{5,6}:
\begin{equation}\label{3.8}
 P^M_{\pm}=\int P_{\pm}(x)\,\rho_x(x)\,dx,
\end{equation}
where $\rho_x(x)$ is related with $\rho_m(m)$ by the standard equality $\rho_x(x)dx=\rho_m(m)dm$. Here, we should note that the values $\Delta m$ and $\Delta\Gamma$ emerge as a result of the diagonalization of the matrix $M-i\Gamma/2$, which describes the mixing in $M^0-\bar{M}^0$ system. So, the values $M_1,\,M_2,\,\Gamma_1,\,\Gamma_2$ and, consequently, $\Delta m$ and $\Delta\Gamma$ are interdependent in the general case. Therefore, variation $x$ leads to variation $y$, that is $y\to y(x)$, where the function $y(x)$ is defined by the functions $\Gamma(x)$ and $\Delta \Gamma(x)$. This issue will be considered in the next section.

The simplest characteristic of mixing is the time-integrated mixing probability:
\begin{equation}\label{3.9}
 \chi(x)=\frac{P_{-}(x)}{P_{-}(x)+P_{+}(x)}\approx\frac{x^2+y^2}{2(1+x^2)},\,\,\,
 (|\frac{q}{p}|^2\approx 1).
\end{equation}
The model generalization of this characteristic is the quantity
\begin{equation}\label{3.10}
 \chi^M=\frac{P^M_{-}}{P^M_{-}+P^M_{+}}=\frac{\int P_{-}(x)\,\rho(x)\,dx}{\int[P_{-}(x)+P_{+}(x)]\rho(x)\,dx}.
\end{equation}
We should note that in the general case $\chi^M\neq \overline{\chi(x)}$. It is easy to check that in the limit of the fixed mass, that is when $\rho(x)=\delta(x-x_0)$, the value $\chi^M$ coincides with the standard one:
\begin{equation}\label{3.11}
 \chi^M=\chi(x_0)=\frac{x^2_0+y^2}{2(1+x^2_0)}.
\end{equation}

Thus, to calculate the contribution of the MSE to the value $\chi$, that is to $\Delta m$, we need the function $\rho_x(x)$ which is defined by the initial function $\rho_m(m)$. The various definitions of  $\rho_m(m)$ were discussed in Refs.\cite{5,6}, where the Lorentzian (Breit-Wigner type), Gaussian and phenomenological distributions have been considered. It was noted, that the Lorentzian and Breit-Wigner distributions have a bad behavior at the infinity. So, for the evaluation of the mass-smearing contribution to the mixing, further we use the Gaussian distribution.

\section{Mass-smearing effect in the $K^0-\bar{K}^0$ system}

In the case of $K^0-\bar{K}^0$ system, the time-integrated measurements are not suitable because of the lifetime $\tau_L$ of the long-lived component $K^0_L$ is much greater the lifetime $\tau_S$ of the short-lived component $K^0_S$, that is $\Gamma_S\gg\Gamma_L$ and $y=(\Gamma_s-\Gamma_L)/(\Gamma_S+\Gamma_L)\approx 1$. From Eq.(\ref{3.7}), it follows that the second term in $P_{\pm}(x)$ is small and $\chi^M\approx\chi\approx 1/2$. So, for this case, the time-dependent characteristics of mixing are usually measured. However, we firstly consider the time-intergated characteristic $\chi$ in order to illustrate the influence of mass smearing on mass splitting. Moreover, the comparison of the standard and model approaches gives us some information on the function $y(x)$.

Consider the model value $\chi^M$ which is defined by the expression (\ref{3.10}). To evaluate $\chi^M$ we need the probability density $\rho(x)$, which follows from the function $\rho(m_S)$ in the case of $K^0-\bar{K}^0$ system. The value $\Delta m_0=M_L-M_S$ is fixed mass difference, defined by FCNC transitions according to $\Delta m_0\approx2\mathit{M}_{12}$. Then, $\Delta m=M_L-m_S$ is a random value of the mass difference, which is defined by the probability density $\rho(m_S)$ with mean value $\overline{m_S}=M_S$ and deviation $\sigma(m_S)=\Gamma_S/2$. Further, we consider Gaussian distribution for the short-lived component:
\begin{align}\label{4.1}
 \rho(m_S)\,dm_S=&\frac{1}{\sqrt{2\pi}\sigma}\exp\{-\frac{(m_S-M_S)^2}{2\sigma^2}\}
 \,dm_S\notag\\
               &=-\frac{1}{\sqrt{2\pi}\sigma}\exp\{-\frac{(\Delta m-\Delta m_0)^2}{2\sigma^2}\}\,d\Delta m,
\end{align}
where $\Delta m=M_L-m_S$, $\Delta m_0=M_L-M_S$, $\sigma=\Gamma_S/2\approx \Gamma$ and $\Gamma=(\Gamma_S+\Gamma_L)/2$. Eq.(\ref{4.1}) can be represented in the form:
\begin{equation}\label{4.2}
 \rho_x(x)\,dx=\frac{1}{\sqrt{2\pi}}\,\exp\{-\frac{1}{2}(x-x_0)^2\}\,dx,
\end{equation}
where $x=\Delta m/\Gamma$, $x_0=\Delta m_0/\Gamma$ and $dx=-d\Delta m/\Gamma$. Now, we should take into account the interdependence of the variables $\Delta m$ and $\Delta \Gamma$. Such a dependence arises in self-consistent diagonalization of generalized mass-decay matrix $M-i\Gamma/2$ of $K^0-\bar{K}^0$ system with variable (random) matrix elements. Moreover, some measurements of the $\tau_S$ (that is $\Gamma_S$) are correlated with $\Delta m$ \cite{16b}. We have no a rigid theory of such a system and define the interdependence of the variable values $\Delta m$ and $\Delta \Gamma$ in a phenomenological way as function $\Gamma_S(x)$. In the simplest case, one can approximate this function by linear relation $\Gamma_S(x)=\Gamma_S+k(x_0-x)$ with free parameter $k$. So, we get an equality $\Gamma_S(x_0)=\Gamma_S$, where $\Gamma_S$ is experimental mean value of width of the short-lived state $K^0_S$. In order to avoid the negative values of the $\Gamma_S(x)$, when $k(x-x_0)>\Gamma_S$, we use the exponential formula:
\begin{equation}\label{4.3}
\Gamma_S(x)=\Gamma_S\exp(k(x_0-x)).
\end{equation}
Expanding this expression in the vicinity of $x_0$, one can see the satisfaction of the above mentioned requirements. Note that the gaussian distribution (\ref{4.2}) cuts out the vicinity of $x_0$ and approximately equates the linear and exponential expressions for $\Gamma_S(x)$.
Now, we have to modify the Eq.(\ref{3.7}), introducing the dependence
$y(x)=2\Delta \Gamma(x)/\Gamma(x)$, where $\Gamma(x)=(\Gamma_S(x)+\Gamma_L)/2$. With allowance for a such modification, Eq.(\ref{3.10}) can be represent in the form:
\begin{equation}\label{4.4}
\chi^M\approx\frac{\int\{1-\epsilon(x)-4\epsilon(x)/(1+x^2)\}\rho(x)dx}{2\int(1-\epsilon(x))\rho(x)dx},
\end{equation}
where $\rho(x)$ is defined by Eq.(\ref{4.2}), $\epsilon(x)=\epsilon\exp(-k(x_0-x))$ and $\epsilon=\Gamma_L/\Gamma_S\sim 10^{-3}$ is small parameter. To illustrate the contribution of mass smearing to mass splitting we introduce the effective (model) $\Delta m^M$ and model parameter $x^M=\Delta m^M/\Gamma$. These parameters are related with the characteristic $\chi^M$ in a standard way (\ref{3.11}):
\begin{equation}\label{4.5}
\chi^M=\frac{(x^M)^2+y^2}{2(1+(x^M)^2)},\,\,\,x^M=\sqrt{\frac{2\chi^M-y^2}
{1-2\chi^M}}\,,
\end{equation}
where $y=y^{exp}$. So, in order to illustrate effect, we assume an additional convention. With the help of Eqs.(\ref{4.4}) and (\ref{4.5}) we get a set of the curves (Fig.1), which describe the dependence $x^M$ on free parameter $k$ for various $\Delta m_0=(1.87,\,2.34,\,2.72)\cdot10^{-6}\,\,\mbox{eV}$, which correspond to $x_0=0.51,\,0.64,\,0.74$  (dashed, dashed and solid lines respectively). The first two values $\Delta m$ have been considered in the second section as short-distance (SD) mass splitting and the third one is found from the condition $x^M(k,\Delta m_0)=x^{exp}=0.9478$.
%-------------------------------------------------------------
\begin{figure}[h!]
\centerline{\epsfig{file=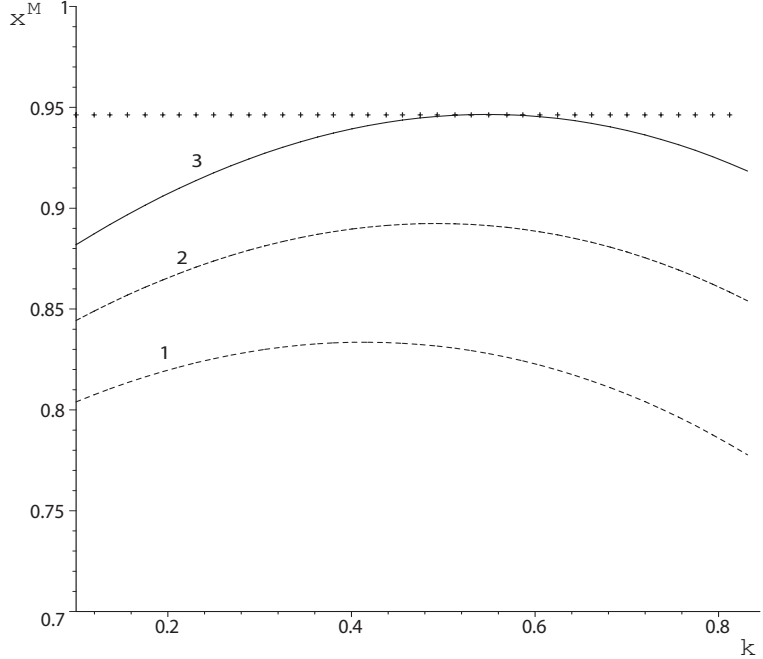,height=9cm,width=10cm}}
\caption{\footnotesize The dependence $x^M$ on parameter $k$ at
various $x_0$ ($\Delta m_0$). (1) $x_0$=0.51;\,(2)$x_0$=0.64;\,(3) $x_0$=0.74.} \label{Kx}
\end{figure}
%-------------------------------------------------------------

From Fig.1, it follows that the mass smearing gives an additional contribution to mass splitting, however, the standard SD mass splitting $\Delta m_0=(1.87;\,\,2.34)\cdot10^{-6}\,\,\mbox{eV}$ is not sufficient to explain the experimental one even with the contribution of mass smearing. The value $\Delta m_0=2.72\cdot10^{-6}\,\mbox{eV}$ is found by the matchings $\Delta m^M$ and $\Delta m^{exp}$ at $k\approx0.55$ (see Fig.1). This results are based on rather rough evaluations which contain some arbitrariness. However, they lead to the conclusions: mass smearing gives an additional contribution to the mass splitting in $K^0-\bar{K}^0$ system; we need in extra small splitting, caused, for instance, by the long distance contribution. Besides, the phenomenological function $\Gamma_S(x)$ can be described by the function (\ref{4.3}) with the parameter $k\sim 0.5$.

Now, we consider in detail the time-dependent effect, which describes the oscillation of strange in the beam of $K$-mesons. The experimental measurements of the $\Delta m$ and $\Gamma_S$ is connected with the complicated procedure of fitting. Experimental data are fit to the distribution of decays in the regenerator beam, where $\Delta m$ and $\Gamma$ are fitting parameters along with other technical parameters \cite{25}. So, we consider the simplest standard characteristic of this effect \cite{26}:
\begin{equation}\label{4.6}
 S(t,\Delta m)=\frac{P_+(t,\Delta m)-P_-(t,\Delta m)}{P_+(t,\Delta m)+P_-(t,\Delta m)},
\end{equation}
where $P_{\pm}(t,\Delta m)$ are defined by Eqs.(\ref{3.4}) and (\ref{3.5}). The characteristic $S(t,\Delta m)$ is convenient to describe the oscillation of strange in $K^0$-meson beam at the time $t\sim \tau_S$. The time of oscillation depends on the splitting of mass, which is smeared within the framework of our model. So, we introduce the weighted probability transition:
\begin{equation}\label{4.7}
 P^M_{\pm}(t)=\int P_{\pm}(t,\Delta m)\rho(\Delta m) d\Delta m,
\end{equation}
where $\rho(\Delta m)$ is probability density (\ref{4.1}). Then, the weighted time-dependent characteristic of oscillation is defined as:
\begin{equation}\label{4.8}
 S^M(t)=\frac{P^M_+(t)-P^M_-(t)}{P^M_+(t)+P^M_-(t)}.
\end{equation}

Performing the change of variable $\Delta m \to x\Gamma(x)$ in the expressions (\ref{4.7}), (\ref{4.8}) and substituting $\rho(x)$, defined by (\ref{4.2}), into (\ref{4.7}), we get the final expression for the function $S^M(t)$, This function describes the time-oscillation of strange with an account of mass smearing
and can be represented in the form:
\begin{equation}\label{4.9}
S^M(t)=2\frac{\int \exp(-\Gamma(x)t-0.5(x-x_0)^2)\cos(x\Gamma_S(x)t/2)dx}
{\int [\exp(-\Gamma_S(x)t)+\exp(-\Gamma_Lt)]\exp(-0.5(x-x_0)^2)dx},
\end{equation}
where $x=2\Delta m/\Gamma_S(x)\approx \Delta m/\Gamma(x)$. With the  help of the expression (\ref{4.9}), we have performed a fitting of the standard oscillation curve $S^{exp}(t)$, which is defined by (\ref{4.6}) and corresponds to the experimental values $\Delta m^{exp}$ and $\Gamma^{exp}$. The input parameter $\Delta m_0$ was changed in the interval $(2.34-2.72)\cdot 10^{-6}\,\mbox{eV}$ and free parameter was varied near the value $k\approx 0.5$. It was found, that the least deviation of the $S^M(t,\Delta m_0)$ from the $S(t,\Delta m^{exp})$ takes place for the values $\Delta m^M_0=\Delta m_0=2.6\cdot 10^{-6}\,\mbox{eV}$ and $k=0.42$.
In Fig.2 we represent the functions $S(t)$ for the cases, when mass differences are equal to $\Delta m^{exp}=3.483\cdot10^{-6}\mbox{eV}$ (dotted line) and $\Delta m_0=2.34\cdot10^{-6}\mbox{eV}$ (dashed line), and the model function $S^M(t)$ for the $\Delta m_0=2.6\cdot10^{-6}\mbox{eV}$ (solid line). The true (proper) time is represented in the unit $6.58\cdot 10^{-10}\,\mbox{sec}$.
%-------------------------------------------------------------
\begin{figure}[h!]
\centerline{\epsfig{file=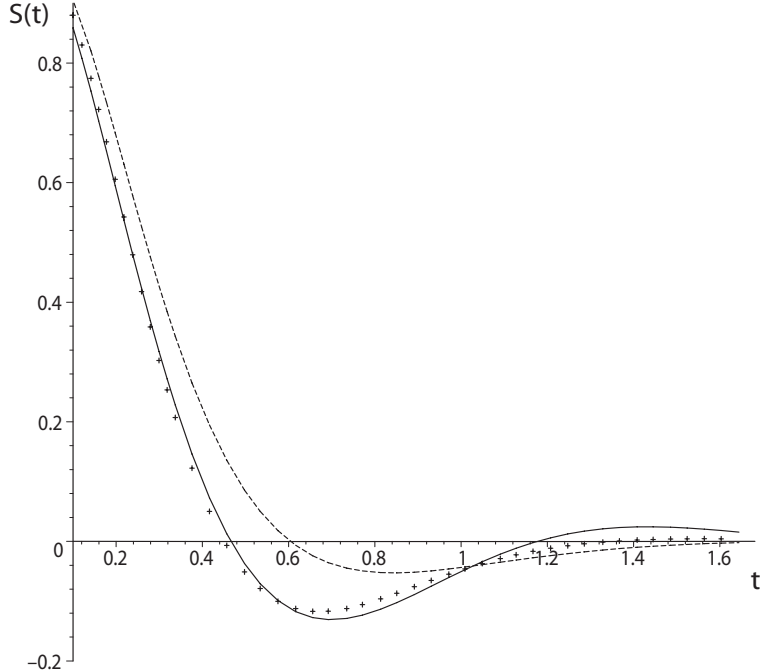,height=9cm,width=10cm}}
\caption{The oscillation of strange in $K$-beam at various $\Delta
m_0$. Dotted line - $S(t,\Delta m^{exp})$; Solid line - $S^M(t,\Delta^M_0)$. True time is $t\cdot 6.58\cdot10^{-10}\,\mbox{sec}$.} \label{St}
\end{figure}
%-------------------------------------------------------------

From Fig.2, one can see that an account of the mass smearing make it possible to decrease the discrepancy between the  theoretical prediction $\Delta m$ and the experimental value $\Delta m^{exp}$. The approach considered can not describe strictly an effect of mass smearing in
the systems of neutral mesons, however, it provides an estimation of this effect.
From the results of estimation, it follows that the effect is large for the case of $K^0-\bar{K}^0$ system and should be accounted in the interpretation of the experimental data. Moreover, this estimation gives an additional information about possible value of the long-distance or new physics contribution.

The above considered approach to the description of the MSE in $K^0$-meson system has rather methodological than an operational status. The time-integrated characteristic $\chi(x)$ is not well defined for this system and can not be experimentally measured. The time-depended characteristic $S(t)$, which describes an oscillation of strange, is more adequate and conventional one. However, time-energy uncertainty relation imposes some restriction on the measuring of this characteristic. To get more detailed function $S(t)$ we have to decrease an interval of time $t+\delta t$, where the measurement is fulfilled. According to uncertainty relation for unstable system the energy, and consequently the mass of $K^0_S$ meson, is measured with the corresponding uncertainty. The oscillation of strange manifests itself at time $t\sim O(\tau_S)$, so the measurements at the intervals $\delta t\lesssim \tau_S$ lead to the large uncertainty. In order to escape this uncertainty, we need in large statistic or another characteristic, for instance, partially time-integrated one, which is some modification of the $\chi(x)$, defined by (\ref{3.9}):
\begin{equation}\label{4.10}
\chi(T,x)=\frac{\int_{0}^{T}P_-(t,x)dt}{\int_{0}^{T}(P_-(t,x)+P_+(t,x))dt}.
\end{equation}
Performing the integration in (\ref{4.10}), we can represent $\chi(T,x)$ in the form:
\begin{equation}\label{4.11}
\chi(T,x)=\frac{1}{2}[1-\frac{I_2(T,x)}{I_1(T,y)}],
\end{equation}
where
\begin{align}\label{4.12}
&I_1(T,y)=\frac{1}{2\Gamma(1-y^2)}[1-e^{-\Gamma T}\cosh(\Delta\Gamma T/2)-ye^{-\Gamma T}\sinh(\Delta\Gamma T/2)],\notag\\
&I_2(T,x)=\frac{1}{2\Gamma(1+x^2)}[1-e^{-\Gamma T}\cos(\Delta m T)+xe^{-\Gamma T}\sin(\Delta m T)].
\end{align}
Here, $x=\Delta m/\Gamma$, $y=\Delta\Gamma/2\Gamma$ and $y\neq y(x)$ when the mass smearing is absent. It is easy to check that in the limit $T\to\infty$, the value $\chi(T,x)$ go to usual function $\chi(x)$ which is defined by Eq.(\ref{3.9}). The measurement of the partially time-integrated characteristic $\chi(T)$ is also in conflict with uncertainty relation at $T\lesssim \tau_S$, but it seems lead to smaller uncertainty in comparison with $S(t)$.

In full analogy with the definition of $S^M(t)$ the model definition of $\chi^M(T)$, which account mass smearing, can be represented by the expression:
\begin{equation}\label{4.13}
\chi^M(T)=\frac{1}{2}[1-\frac{I^M_2(T)}{I^M_1(T)}],
\end{equation}
where
\begin{equation}\label{4.14}
I^M_{1,2}(T)=\int I_{1,2}(T,x)\rho(x)dx,
\end{equation}
and $x=\Delta m/\Gamma(x)$, $y(x)\approx 1-2\epsilon(x)$.  In Fig.3 we represent the functions $\chi(T)$ (dotted and dashed lines) and $\chi^M(T)$ (solid line) for the same condition as in Fig.2. Thus, the approach considered makes it possible to improve an accordance between the standard evaluations of the mixing parameters and experimental data on the oscillation in $K^0-\bar{K}^0$-system. Moreover, we establish a correspondence between the time-integrated characteristics of the mixing and time oscillation of strange.

%-------------------------------------------------------------
\begin{figure}[h!]
\centerline{\epsfig{file=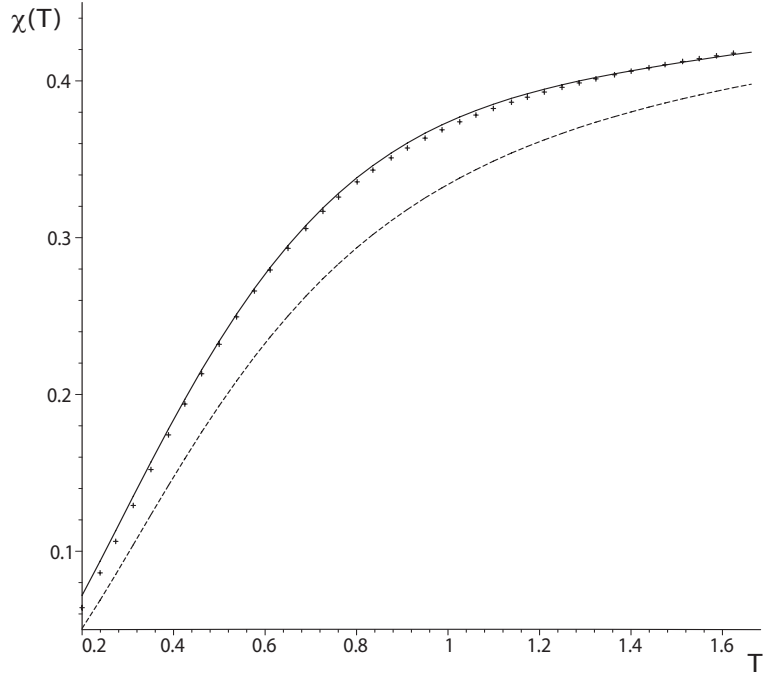,height=9cm,width=10cm}}
\caption{The time-integrated characteristic of oscillation in
$K$-beam at various $x_0$ ($\Delta m_0$). Dotted line - $\chi(T,x^{exp}$); Solid line - $\chi^M(T,x^M_0)$. True time is $T\cdot 6.58\cdot10^{-10}\,\mbox{sec}$.} \label{chit}
\end{figure}
%-------------------------------------------------------------

 We have performed an estimation of the effect of mass smearing in the $K^0-\bar{K}^0$ system. This estimation has some arbitrariness related with free parameter $k$ in the function $y(x,k)$ and with the choice of the $\Delta m_0$. These free input are fixed by the matching of the model characteristic of oscillation with the standard one, when the experimental $\Delta m$ is used. The approach considered make it possible to perform this matching with good accuracy and evaluate the MSE, which occur noticeable.

\section{Mass-smearing effects in $B$ and $D$ systems}

The time-integrated value $\chi$, defined by Eq.(\ref{3.9}), is well measured characteristic of mixing in the case of $B^0_{d,s}-\bar{B}^0_{d,s}$ and $D^0-\bar{D}^0$ systems. In these cases, we have to take into consideration an inequality $\Delta \Gamma\ll\Gamma$, that is $\Gamma_L\approx\Gamma_H$. So, both the states $M^0_L$ (light) and $M^0_H$ (heavy) have approximately the same width $\Gamma$ and, consequently, the same mass smearing. To describe mass smearing by the probability density $\rho(\Delta m)$ we need in two-dimensional function $\rho(m_H,m_L)$, which in the general case has five parameters - $\bar{m}_H,\,\bar{m}_L,\,\sigma_H,\,\sigma_L$ and $\rho_k$ (coefficient of correlation). The values $\bar{m}_H=M_H$ and $\bar{m}_L=M_L$ are related by the equality $\bar{m}_H-\bar{m}_L=M_H-M_L=\overline{\Delta m}$. From the relation $\Gamma_H\approx\Gamma_L=\Gamma$ it follows $\sigma_H\approx\sigma_L=\sigma$, where $\sigma\approx\Gamma/2$. So, we have one free parameter $\rho_k$. From the physical consideration, it is clear that for $\overline{\Delta m}\gg\Gamma$, which occur in the $B^0_S-\bar{B}^0_S$ system, the correlation is small, $\rho_k\approx0$. When $\overline{\Delta m}/\Gamma=\bar{x}$ decreases then the correlation increases. So, it is naturally to assume that for the $B^0_d-\bar{B}^0_d$ system ($x^{exp}_d=0.776$) the correlation is not large (moderate) and for the $D^0-\bar{D}^0$ system ($x\approx10^{-2}$) it is large.

The probability density $\rho(\Delta m)$ can be exactly defined with the help of the assumption that the conditional one-dimensional distribution of values $m_H$ and $m_L$ is Gaussian. Then, in accordance with the comment to the Theorem 18.8-9(b) in Ref.\cite{27}, the value $\Delta m=m_H-m_L$ is also described by Gaussian distribution with the parameters:
\begin{align}\label{5.1}
&\overline{\Delta m}=\bar{m}_H-\bar{m}_L=\Delta m_0=M_H-M_L;\notag\\
&\sigma^2(m_H-m_L)=\sigma^2(m_H)+\sigma^2(m_L)-2\rho_k\sigma(m_H)\sigma(m_L).
\end{align}
Taking into account the equalities $\sigma(m_H)=\sigma(m_L)=\Gamma/2$ and $\sigma(\Delta m)=\Gamma\sqrt{(1-\rho_k)/2}$, we represent the differential of probability in the form:
\begin{equation}\label{5.2}
\rho(\Delta m)d\Delta m\rightarrow \rho_x(x)dx=\frac{1}{\sqrt{\pi(1-\rho_k)}}\exp(-\frac{(x-x_0)^2}{1-\rho_k})dx,
\end{equation}
where $x=\Delta m/\Gamma$ and $x_0=\Delta m_0/\Gamma$. In the limit $\rho_k\to 1$ (full correlation) the function $\rho_x(x)\to \delta(x-x_0)$, that is we get an asymptotic model of $\delta$-function (see \cite{27}, sect. 21.9-4). In this limit, mass difference is fixed and mass smearing does not get any contribution to the mass splitting. So, we consider $0\leq\rho_k <1$, taking into account that maximum effect occurs at $\rho_k=0$, when $\sigma(\Delta m)=\sqrt{2}\sigma(m)$. It is clear also, that in the case $\Delta \Gamma/2\Gamma=y\ll1$, we can neglect the dependence $y=y(x)$. Thus, from Eqs.(\ref{3.7}) and (\ref{3.10}) it follows:
\begin{equation}\label{5.3}
\chi^M=\frac{1}{2}[1-(1-y^2)\int\frac{\rho(x)dx}{1+x^2}],
\end{equation}
where $\rho(x)$ is defined by Eq.(\ref{5.2}).

Now, we consider an effect of the mass splitting in $B^0_d-\bar{B}^0_d$ system. As was shown in Section II, the mass difference in the framework of the Standard Model gives the most likely interval of values $\Delta m^{box}_d\approx(3.52-3.65)10^{-4}\,\mbox{eV}$ ($x_0=0.82-0.85$). These values slightly exceed the experimental one $\Delta m^{exp}_d=(3.337\pm0.033)\cdot10^{-4}\,\mbox{eV}$ ($x_d=0.768-0.784$). An account of the MSE permits to decrease this discrepancy. In Fig.4, we represent the model values $x^M(x_0,\rho_k)$, defined by the relation (\ref{4.5}) with $y=0$, as the functions of $\rho_k$ for $x_0=0.79,\,0.80,\,0.81$ (dashed, solid and dashed lines, respectively).
%-------------------------------------------------------------
\begin{figure}[h!]
\centerline{\epsfig{file=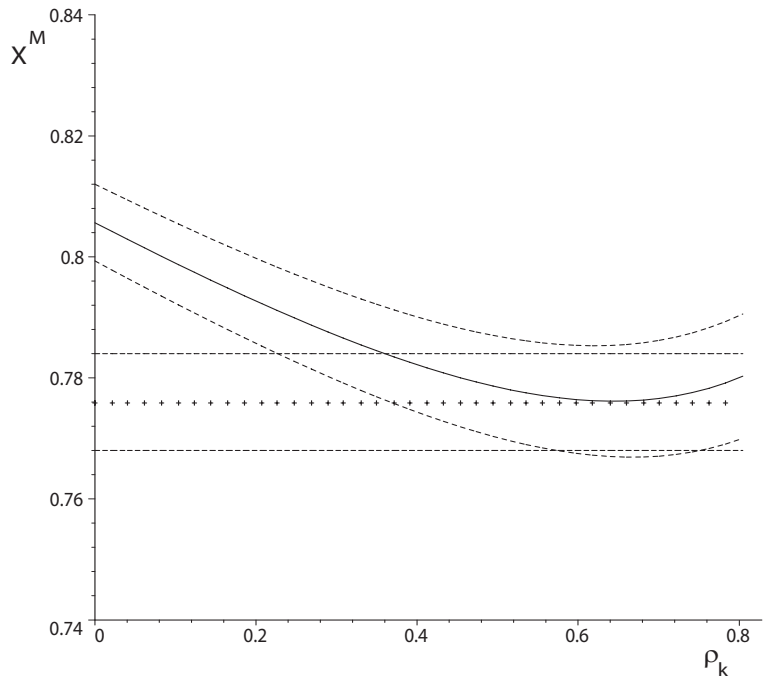,height=9cm,width=10cm}}
\caption{The values $x^M$ as function of $\rho_k$ at
$x_0=0.79,\,0.80,\,0.81$. Solid line - $x^M(\rho_k,x^M_0=0.80)$.} \label{xrb}
\end{figure}
%-------------------------------------------------------------
The horizontal lines restrict the experimental interval of $x$, where the point-like line corresponds to the central experimental value $x=0.776$. The nearest line to the central experimental one corresponds to $\Delta m_0=3.45\cdot10^{-4}\,\mbox{eV}$, which corresponds to $x_0=0.80$, and $\rho_k=0.62$ (moderate correlation). So, an account of the mass smearing make it possible to explain some exceeding of the expected theoretical $\Delta m_0$ over the experimental one. Note that, in contrast to the case of $K^0$-meson, when mass smearing increases mass splitting, in the case of $B^0_d$-meson it decreases mass splitting. To illustrate the effect we calculate the value $R=x^M/x_0$ (coefficient of smearing) as function of $x_0$ at a various $\rho_k$. The results are represented in Fig.5, where $R(x_0)$ are calculated for $\rho_k=0,\,0.5,\,0.8,\,0.99$ (the larger $\rho_k$ the nearer the curve to $R=1$).
%-------------------------------------------------------------
\begin{figure}[h!]
\centerline{\epsfig{file=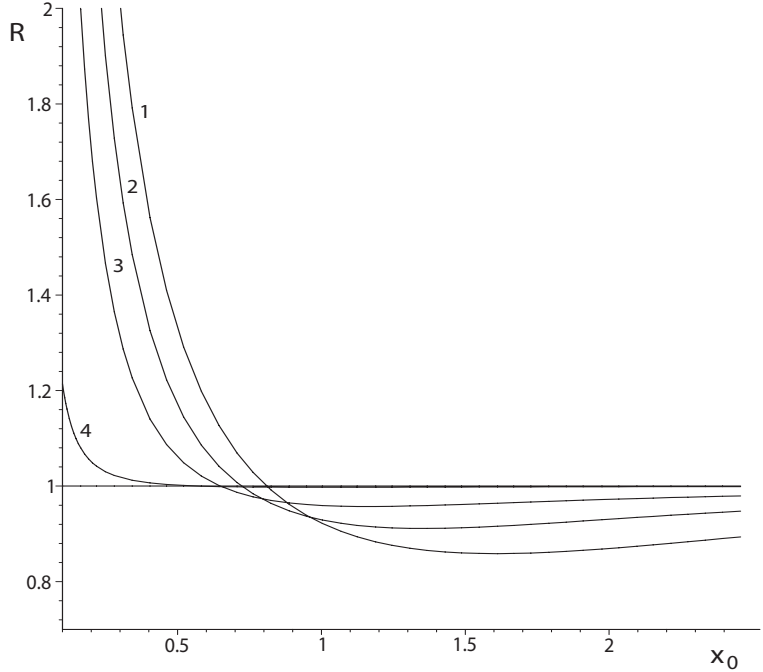,height=9cm,width=10cm}}
\caption{The coefficient of smearing $R=x^M/x_0$ as function on
$x_0$ at various $\rho_k$: (1) $\rho_k$=0;\,(2)$\rho_k$=0.5;\, (3) $\rho_k$=0.8;\, (4) $\rho_k$=0.99.} \label{Rx0}
\end{figure}
%-------------------------------------------------------------
From Fig.5, one can see that for every $\rho_k$ there are two domain of $x_0$, which correspond to $R>1$ and $R<1$. For small $x_0$ the coefficient $R>1$ and mass smearing dominates, when $\rho_k\neq 1$. For $x_0\gtrsim1$ the coefficient $R<1$ and $R(x_0)\to1$ when $x_0\to\infty$, or $\rho_k\to1$. These properties illustrate the role of the MSE in $K^0,\,B^0$ and $D^0$ meson systems. From Fig.5 it follows, that this effect is negligible in $B^0_s$ system and can be very large in $D^0$ system.

In the second section, it was noted that for $D^0-\bar{D}^0$ system the values $x^{th}_D$ and $y^{th}_D$ are expected to be well below $10^{-2}$ in the frame-work of the Standard Model (see also \cite{28}). At the same time, recent measurements imply $x_D\approx10^{-2}$ and $y_D\sim 10^{-3}-10^{-2}$ \cite{9}. An account of the MSE makes it possible to explain this contradiction. When $\Delta m\ll\Gamma$, the main contribution to the mass difference is caused by the mass smearing. From Eq.(\ref{5.3}), it follows that, when $y\ll x\ll 1$, the value $\chi^M$ can be approximated by the formula:
\begin{equation}\label{5.4}
\chi^M(x_0)\approx\frac{1}{2}[1-\int(1-x^2)\rho(x)dx]=\frac{1}{2}(x^2_0+\sigma^2_x),
\end{equation}
where $\sigma^2_x=(1-\rho_k)/2$. Thus, for $\rho_k=0$ and $x_0\ll 1$ we get $\chi^M\approx 1/4$, and according to Eq.(\ref{4.5}), $x^M\approx\sqrt(2\chi^M/(1-2\chi^M))\approx1\gg x^{exp}$. So, the measured value $x^{exp}\sim10^{-2}$ can be obtained in the case of large correlation $\rho_k\approx1$. From Eq.(\ref{5.4}) and $\chi^M=\chi^{exp}$ it follows
\begin{equation}\label{5.5}
\rho_k\approx 1-2(x^{exp})^2.
\end{equation}
Thus, to explain the mixing $x_D\sim10^{-2}$ we have to assume large correlation $\rho_k\approx 1$. As was noted early, this assumption is natural for the case $\Delta m\ll\Gamma$. It should be noted, also, that an account of the dependence $y=y(x)$ may influence on the effective value of the parameter $y$.

\section{Conclusion}

The FCNC processes play a significant role in the precise test of SM and search for New Physics (NP). Mass difference between light and heavy component is an important characteristic of the mixing in $M^0-\bar{M}^0$ systems, which is caused by FCNC at loop level in the frame-work of SM. There are some discrepancy between theoretical predictions on $\Delta m$ and experimental ones. The largest contradiction occur in the case of $D^0-\bar{D}^0$ system ($\Delta m^{exp}\gg\Delta m^{th}$) and $K^0-\bar{K}^0$ one ($\Delta m^{exp}\approx 2\Delta m^{th}$). These facts are often interpreted as the presence of non-controlled processes, for instance LD contribution, or as the windows for NP.

In this work, we show that the phenomenon of mass smearing (or finite-width effect) gives a significant contribution to mass difference in the $K^0$ and $D^0$ meson systems. In particular, this contribution can be dominant in the case of $D^0$ meson and large in the case of $K^0$-meson. An account of the MSE can significantly improves an accordance of the theoretical predictions and the experimental data on $\Delta m$ for $K^0$, $B^0_d$ and $D^0$ mesons. We also suggest the existing of the correlation between the variable masses of the light and heavy states. The value of this correlation depends on the relation of $\overline{\Delta m}$ and $\Gamma$, that is on the value of $\bar{x}$.

The approach considered is the phenomenological description of the mass-smearing phenomenon in the systems of neutral mesons. It based on the model of unstable particles with a smeared mass and closely connected with the uncertainty relation. This effect may occur in other characteristics of mixing, such as width difference $\Delta\Gamma$ (see the third section) or CP violation. Moreover, it may be observed also in the mixing of hadron states with $M_1-M_2\sim\Gamma$, such as $\omega-\rho^0$, $a_0-f_0$ etc. There is no a strict theory of the mass-smearing phenomenon, however, the effect can be described in a phenomenological way and it should be taken into consideration in the precision test of SM and in the search for the NP.

\end{document}